\documentclass[
    a4paper, 
    amsfonts, 
    amssymb, 
    amsmath, 
    reprint, 
    showkeys, 
    footinbib, 
    longbibliography, 
    twoside,
    notitlepage,
    aps,
]{revtex4-2}

\usepackage{graphicx}
\usepackage{dcolumn}
\usepackage{bm}

\newcommand{\blue}[1]{{\color{black}#1}}

\newcommand{\iu}{\mathrm{i}\mkern1mu}
\newcommand{\eu}{\mathrm{e}\mkern1mu}
\let\vec=\mathbf

\usepackage[normalem]{ulem}
\usepackage{multirow}
\usepackage{amsfonts}
\usepackage{physics}
\usepackage{comment}
\usepackage{amsthm}
\usepackage{mathtools}
\usepackage{physics}
\usepackage{xcolor}
\usepackage[left=23mm,right=13mm,top=35mm,columnsep=15pt]{geometry} 
\usepackage{adjustbox}
\usepackage{placeins}
\usepackage[T1,T2A]{fontenc}
\usepackage{csquotes}
\usepackage{float}

\usepackage{multirow}

\usepackage{blindtext}

\usepackage{bm}

\usepackage{xcolor}

\definecolor{urlcolor}{HTML}{3333B2}
\definecolor{citecolor}{HTML}{3333B2}
\definecolor{black}{HTML}{000000}

\usepackage{xcolor}

\usepackage{graphicx}

\graphicspath{}
\DeclareGraphicsExtensions{.pdf,.png,.jpg}

\usepackage{array}
\usepackage{placeins}
\usepackage{tcolorbox}

\usepackage[english]{babel}
\usepackage{comment}
\usepackage[pdftex, pdftitle={Article}, pdfauthor={Author}]{hyperref} 
\hypersetup{citecolor=citecolor, urlcolor=urlcolor, colorlinks=true}
\let\vec=\mathbf

\begin{document}

\preprint{APS/123-QED}

\title{When does nonlinear circular dichroism appear in achiral dielectric nanoparticles?}

\author{Anastasia Nikitina}
\email{anastasia.nikitina@metalab.ifmo.ru}
\author{Anna Nikolaeva}%
\author{Kristina Frizyuk}%
 \email{k.frizyuk@metalab.ifmo.ru}
\affiliation{%
 The School of Physics and Engineering, ITMO University, Saint-Petersburg, Russia \\ 
}%

\date{\today}

\begin{abstract}
Sometimes it appears, sometimes it does not.
We present a theoretical study of \emph{circular dichroism} in the \emph{second-harmonic} signal of \blue{single dielectric nanostructures} with different symmetries and noncentrosymmetric materials.
We show that this effect is not fully identified only by macroscopic and lattice symmetries, but their relative orientation as well. Based on the cumbersome symmetry and modal analysis, we provide a general and very simple formula to determine whether the dichroism exists, and a table with the most significant cases. 

\end{abstract}

\maketitle


\paragraph*{Introduction.} 
Chirality is a naturally occurring property that plays a significant role in physics, chemistry, biology, and medicine. Many examples of chirality can be found in nature, for example, amino acids, which are frequently have an inherent one-handedness~\cite{Pasteur, Cahn1966-SpecificationofMole}. 

\emph{Circular dichroism (CD) spectroscopy} is an excellent method for evaluating the unique properties of proteins, molecules and chiral materials~\cite{Greenfield_2006,Hopkins2016-Circulardichro,DeSilva2021-UsingCircularDichro, Graf2019-AchiralHelicityPre}. {Due to the predominantly weak expression of chiral effects in natural media, chiral metamaterials~\cite{Gorkunov2020-MetasurfaceswithMax, zhao2012twisted,Kim2020-GiantNonlinearCircu}  and chiral plasmonic structures~\cite{PhysRevX.2.031010, li2020tunable, hentschel2017chiral, collins2017chirality,Chen2021-ExtremizeOpticalChi}, \blue{ for example, with Fano-enhanced CD~\cite{Kondratov2016-Extremeopticalchira,Liu2020-Fano-EnhancedCircula}}, have been widely used in recent years as a useful tool for achieving strong chiroptical responses.}

\begin{figure}[ht!]
\centering
\includegraphics[width=0.9\linewidth]{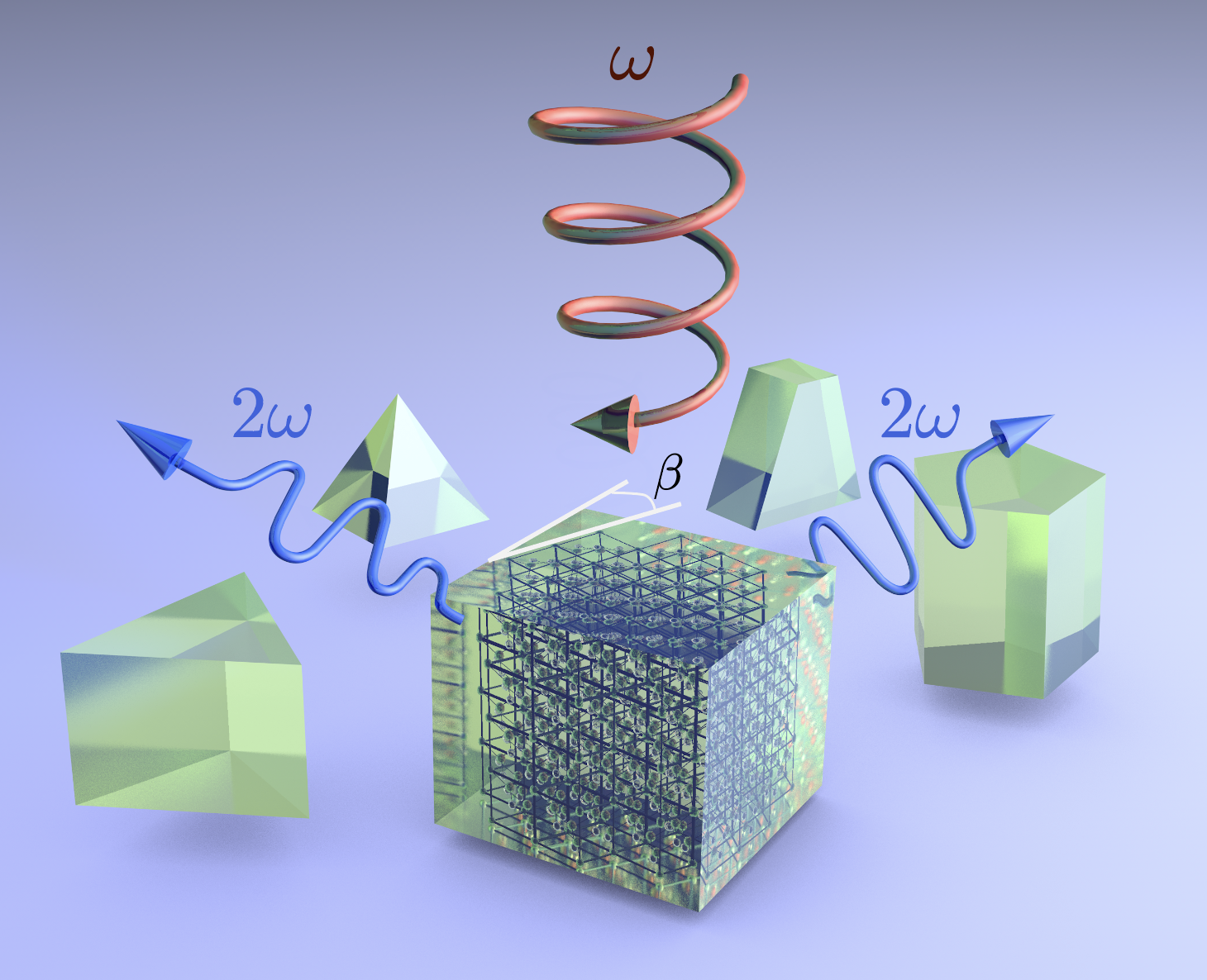}
\caption{Schematic of the concept.  \blue{Separate symmetric  nanoparticle (for example, cube, prism or pyramid)} with a certain lattice symmetry and orientation generates the second harmonic when excited by a circularly polarized wave, where {$\beta$ is the relative angle between the crystalline lattice and nanoparticle}. \blue{Different nanoparticle examples are shown.}}
\label{mainscheme}
\end{figure}

{Circular dichroism can be obtained in nonlinear optical processes, such as second- and third-harmonic generation. Circularly-polarized light interacting with resonances of nanostructures in the nonlinear regime can produce a significant chiroptical signal, which is of great interest due to its high sensitivity to symmetry breaking~\cite{verbiest1994nonlinear,Koshelev2022-Enhancednonlinearci,Weissflog2022-Far-FieldPolarizatio, PhysRevLett.107.257401,mi11020225}.} 
Generally, this nonlinear effect is observed in chiral plasmonic structures, for example, G-shaped structures~\cite{PhysRevLett.104.127401,Valev2009-PlasmonicRatchetWhe,Mamonov2014-Anisotropyversuscir}, structures with C$_\mathfrak{n}$ \blue{rotational symmetry~\cite{Kim2020-GiantNonlinearCircu,Kim2020-DielectricChiralMet,Collins2018-EnantiomorphingChira}, or other chiral asymmetric shapes~\cite{Kang2020-NonlinearChiralMeta, Schreiber2013-ChiralplasmonicDNA,Gandolfi2021-Near-unitythird-harm,Spreyer2022-SecondHarmonicOptic}.} 
Meanwhile, dielectric nanostructures with a huge second-order susceptibility due to the lack of inversion symmetry of their crystalline lattice~\cite{Liu2022-Third-andSecond-Har,Gigli2022-All-dielectricxn,Rocco2022-Secondordernonlinea,carletti2016shaping} are also a promising direction in circular dichroism spectroscopy. Owing to the presence of Mie-resonanes~\cite{Smirnova2018-Multipolarsecond-har,Chen2020-GlobalMieScattering}, anapole states~\cite{Diaz-Escobar2021-Radiationlessanapole,Hu2022-Achievingmaximumsca}, quasi-bound states in the continuum~\cite{Gandolfi:21, Koshelev_2020}, they can generate locally enhanced electromagnetic fields needed for sensitive CD measurements. Various approaches have been already tried~\cite{yoo2015enhancement, doi:10.1021/acs.jpcc.8b10975, ho2017enhancing}.
Sometimes, chiroptical response appears unexpectedly in objects that, at first glance, are not chiral. This can happen due \blue{to the symmetry breaking at arbitrary incidence on the metasurface} ~\cite{Kruk2015-Polarizationproperti, Verbiest1996-OpticalActivityofA,Chen2019-StrongNonlinearOpti,Maoz2012-ChiropticalEffectsi,Plum2009-Metamaterials:Optica,Gompf2011-PeriodicNanostructur}, ``structural'' effects (\blue{relative orientation of the meta-atoms in the array)~\cite{Volkov2009-Opticalactivityind,Moocarme2017-Meta-OpticalChiralit}, or some other nontrivial reason, such as different coupling to the multipole moments of the structure~\cite{Zambrana-Puyalto2014-Angularmomentum-indu}, rotation of the particles~\cite{Pan2019-CircularDichroismin}, asymmetric excitation~\cite{Zu2018-Deep-SubwavelengthRe}, plasmon enhancement of small chiroptical response~\cite{Li2022-Plasmon-MediatedChir}, or even in single atoms via oriented intermediate state~\cite{Ilchen2017-CircularDichroismin}}. These cases are extremely interesting, while for the second harmonic (SH) in achiral nanostructures, strong nonlinear CD can appear, as it was shown in~\cite{Frizyuk2021-NonlinearCircularDi}. In this paper we provide a general rule for determining when nonlinear circular dichroism occurs in achiral dielectric nanoparticles, a very fast and convenient method that will save time for researchers in this rapidly developing field, and could be especially useful for chiral sensing~\cite{Garcia-Guirado2020-EnhancedChiralSensi,Kim2022-Molecularchiralityd,Czajkowski2022-Localversusbulkcir,Both2022-NanophotonicChiralS}. \blue{Indeed, if the nanoparticle or nanostructure~\cite{Yao2018-Enhancingcirculardi,Rui2022-Surface-EnhancedCirc} is used to amplify the chiral signal, the chiral signal from the structure itself must be excluded. In addition, if we observe strong dichroism in the second harmonic, this will also affect the linear response, since a different amount of energy will be converted into a doubled frequency.}
\paragraph*{Results.} 
We analyze the possibility of observing SH-CD in achiral nanoparticles with different symmetries and materials. We consider the following geometry:
the nanoparticle of a symmetry $\text{C}_{\mathfrak{n}\text{v}}$ or $\text{D}_{\mathfrak{n}\text{h}}$ \blue{(pyramid and prism with a regular polygon in the base~\cite{Boyd2003})} with the symmetry axis along the $z$-axis is irradiated by a normally incident circularly polarized plane wave (L/RCP is the Left/Right circular polarization), see Fig.~\ref{mainscheme}. {We assume that a nanoparticle is made by lithography technique from a monocrystalline bulk material. In this case, we can ``cut'' this nanoparticle under different angles with respect to a lattice orientation. We fix lattice orientation, and suppose, that the nanoparticle is rotated by the angle $\beta$, as it is shown also in Fig.~\ref{mainscheme}. The existence of a substrate does not alter the result. 
}
\begin{figure}[ht!]
\centering
\includegraphics[width=0.99\columnwidth]{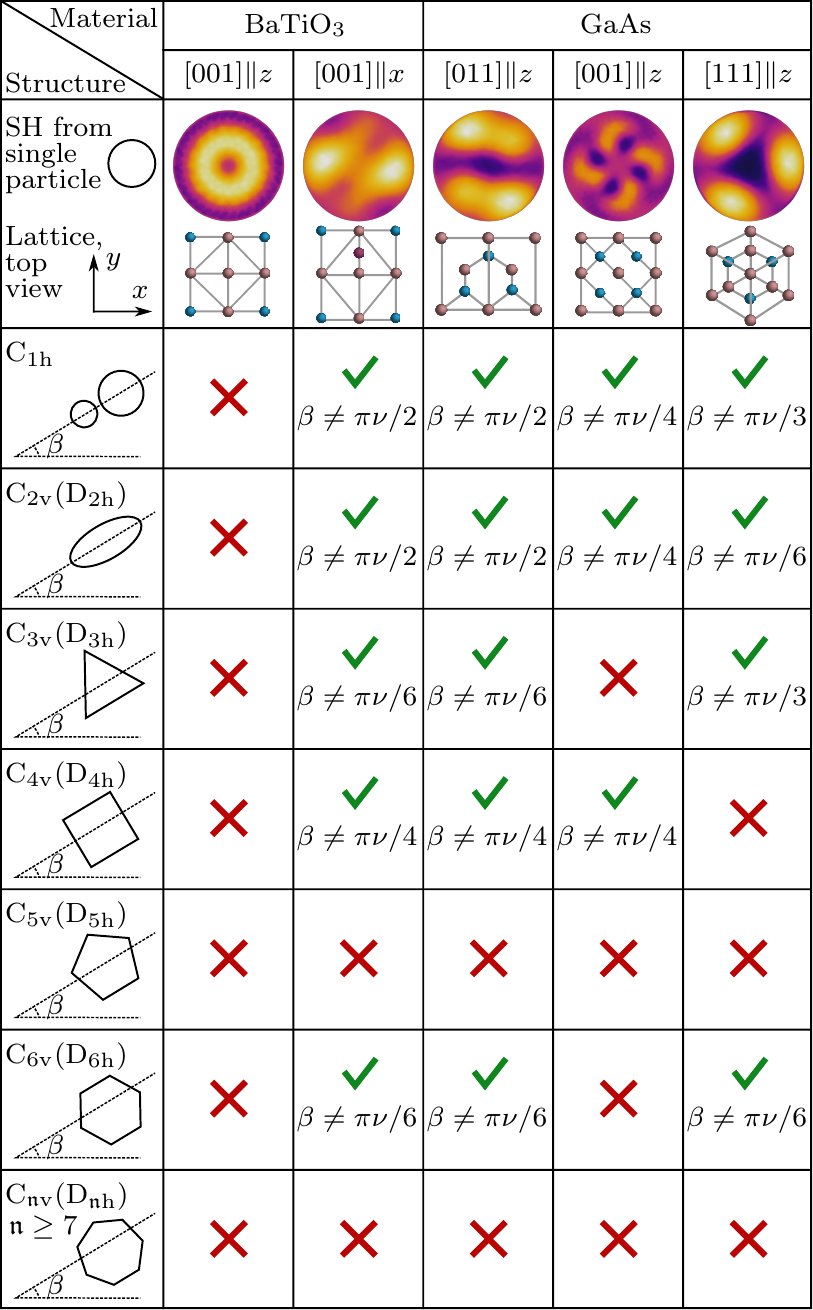}
\caption{The table with main results on the existence of nonlinear circular dichroism in achiral nanoparticles with different symmetries and crystalline lattice. The upper row \blue{shows the angular dependence of the SH power flow into the upper half plane} for a single cylindrical or conical particle for each crystalline lattice, as well as top view of lattices. Below, for different symmetry groups we noted whether SH-CD appears for any angles of nanoparticle rotation $\beta$, except a finite number of angles, or it is not possible at all.\label{mainresults}}
\end{figure}
The SH-CD is given by the following formula~\cite{Bertolotti2015-Secondharmoniccircu,Rodrigues2014-NonlinearImagingand,Byers1994-Second-harmonicgener}: 
\begin{equation}
  \text{SH-CD}= 2\frac{(I_\text{RCP}^{2\omega}-I_\text{LCP}^{2\omega})}{(I_\text{RCP}^{2\omega}+I_\text{LCP}^{2\omega})}
  \label{SHCD}
\end{equation}
where $I^{2\omega}_\text{RCP/LCP}$ --- the total second-harmonic intensity integrated over the angles for RCP and LCP incident waves respectively. In case, if these intensities are different, we say, that nonlinear circular dichroism appears. 

The main result is presented in Fig.~\ref{mainresults}. 
In the left column, the symmetries of a particle (with examples of such shapes) are shown. Let us note, that the exact shape does not play role. For example, we should consider a trimer structure in the same way as a triangular prism, while they both possess $\text{D}_{3\text{h}}$ symmetry. {
The incident light wavelength also does not play a major role, but for a more pronounced SH-CD, it was chosen in the vicinity of first Mie resonances.}

{
In the left column we show some relative angle between the lattice and nanoparticle $\beta$, while the lattice is fixed as shown in the upper row. \blue{Our method is applicable for any crystalline lattice orientation, if the initial (not rotated) nonlinear susceptibility tensor $\hat\chi^{(2)}$ is known. ``Rotated'' tensor  for different orientations is obtained with the Matlab code given in the Supplementary~\cite{gitt}. In our considerations, we assume that all fields are decomposed into a series of multipoles~\cite{Gladyshev_Frizyuk_Bogdanov_2020, Frizyuk_Volkovskaya_Smirnova_Poddubny_Petrov_2019}, but we will be interested only in the total angular momentum projection $m$.} In the upper row we also provide the SH power flow into upper half plane angular distribution for a single cylindrical or conical particle. \blue{Such patterns with several petals can only appear, if the radiation possess several different angular momentum projections $m$. The number of petals is connected with the difference between the projections.} 
\begin{figure}[ht!]
\centering
\includegraphics[width=0.9\linewidth]{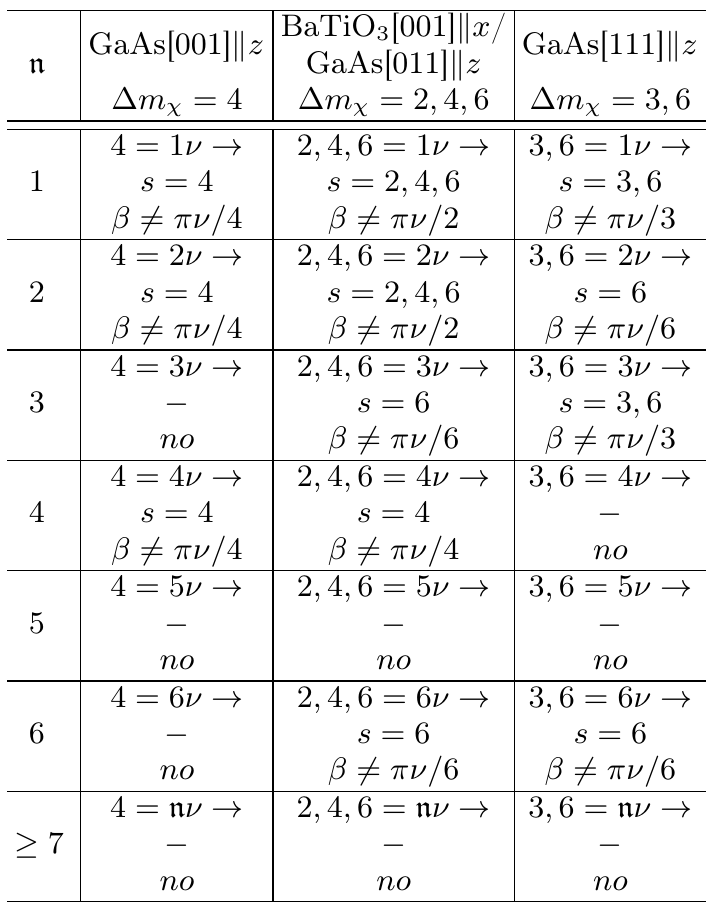}
\caption{The illustration of the main rule for different cases. For each crystalline lattice and symmetry group $\text{C}_{\mathfrak{n}\text{v}} (\text{D}_{\mathfrak{n} \text{h}})$ we show number $s$ and angles $\beta$, when nonlinear circular dichroism exists. If SH-CD is not possible, "no" is written. {In the upper row the difference $\Delta m_\chi$ between the angular momentum projections in tensor $\hat\chi^{(2)}$ is also provided.}}
\label{mainrule}
\end{figure}
{In cells of the Fig.~\ref{mainresults}} we put a \blue{check mark}, if the dichroism appears. It usually exists for any angle $\beta$, except a finite number of angles, which are also given in each cell. We put a cross, if SH-CD never appears for such particle shape and orientation. 


\paragraph*{Theory.}
\blue{Despite the fact that the theory of mode excitation in the second harmonic has already been sufficiently developed in previous works~\cite{Frizyuk_Volkovskaya_Smirnova_Poddubny_Petrov_2019,Frizyuk2021-NonlinearCircularDi}, the consideration of each individual case requires a large amount of time and calculations. Additional complexity is introduced by the fact that usually both in the first and in the second harmonic several modes of different symmetry are excited, and all of them must be considered differently, depending on whether they are degenerate or not. However, we propose a simple version of the consideration, based on the law of conservation of momentum projection~\cite{Frizyuk_2019,Frizyuk_Volkovskaya_Smirnova_Poddubny_Petrov_2019}. It consists of several simple steps:
\begin{enumerate}
    \item {The incident circularly polarized plane wave has a projection $m^{\text{wave}}=\pm 1$. Thus, it excites the modes inside the nanoparticle.}
    \item {If the nanoparticle has $\mathfrak{n}$-fold rotational symmetry, the additional momentum comes, which is the multiple of  $\mathfrak{n}$. Thus, on the fundamental frequency modes containing terms with the following $m$'s are excited inside the nanoparticle: $m^{\text{in}}=\pm 1+\mathfrak{n}\nu $, $\nu \in \mathbb Z$ }
    \item{The nonlinear polarization is written as follows $ \vec P^{2\omega} (\vec r)=\varepsilon_0 \hat\chi^{(2)} \vec E^\text{in} (\vec r)\vec E^\text{in} (\vec r)$, where the field inside the nanostructure $\vec E^\text{in} (\vec r)$ possess the momentum $m^{\text{in}}$, and additional momentum $m_{\chi}$ comes due to the lattice symmetry. It can be easily obtained by rewriting the tensor in cylindrical coordinates~\cite{Toftul2022-Nonlinearity-induced}.  }
    \item{Combining all the additional terms, we obtain that modes with $m^{2\omega}=\pm2+\mathfrak{n}\nu+m_{\chi}$ are excited on the doubled frequency with the additional phase $-m_\chi\beta$, which is also seen from the tensor in cylindrical coordinates.}
    \item{It's worth noting, that there are always several different $m_\chi$. 
    We introduce $\Delta m_\chi$ as the difference between any two $m_\chi$s.
    }
    \item{Consider the excitation of two modes with the difference $\Delta m_\chi$. For LCP, the polarization will be proportional to 
    \begin{equation}
    \begin{aligned}
        \sum_\nu  \eu^{-\iu m_\chi\beta} [\vec P^{2\omega}_{m^{2\omega},\nu}(r,z)\eu^{\iu\varphi(2+\mathfrak{n}\nu+m_{\chi})}+\\+\vec P^{2\omega}_{m^{2\omega}+\Delta m_\chi,\nu}(r,z)\eu^{\iu\varphi(2+\mathfrak{n}\nu+m_{\chi}+\Delta m_\chi)}\eu^{-\iu\Delta m_\chi\beta}]
    \end{aligned}
    \end{equation} 
    For RCP, we consider the modes with opposite $m^{2\omega}$, while due to the tensor structure they are always presented:
       \begin{equation}
    \begin{aligned}
        \sum_\nu \eu^{\iu m_\chi\beta} [\vec P^{2\omega}_{-m^{2\omega},\nu}(r,z)\eu^{\iu \varphi(-2+\mathfrak{n}\nu-m_{\chi})}+\\+\vec P^{2\omega}_{-m^{2\omega}-\Delta m_\chi,\nu}(r,z)\eu^{\iu\varphi(-2+\mathfrak{n}\nu-m_{\chi}-\Delta m_\chi)}\eu^{\iu\Delta m_\chi\beta}]
    \end{aligned}
    \end{equation} }
    \item{Now, we remember that the structure is not chiral, thus, modes with opposite $m^{2\omega}$ are excited with the same weight by the same polarization terms ($|P^{2\omega}_{m^{2\omega},\nu}(r,z)|=|P^{2\omega}_{-m^{2\omega},\nu}(r,z)|$). However, the phase between them, $\eu^{\iu\Delta m_\chi\beta}$, is different for LCP and RCP cases, which means that if these modes interfere, they interfere differently for two incident wave polarizations. So, for the SH-CD to appear, we need these modes to interfere, and $\eu^{\iu\Delta m_\chi\beta}\neq \eu^{-\iu\Delta m_\chi\beta}$.}
\end{enumerate}
}
\paragraph*{Main rule.} 
\blue{Summarizing the above considerations}, we derived the rule, which is applicable for all cases of SH-CD in achiral structures. 
We only should know the susceptibility tensor  $\hat{\chi}^{(2)}$ in cylindrical coordinates 
{
and symmetry group $\text{C}_{\mathfrak{n}\text{v}} (\text{D}_{\mathfrak{n} \text{h}})$ of nanoparticle. Tensor $\hat{\chi}^{(2)}$ in cylindrical coordinates contains the exponential terms, $\eu^{\iu m_\chi \varphi}$, so we introduce $\Delta m_\chi$ as the difference between any two $m_\chi$. For several lattices, all possible $\Delta m_\chi$ are given in the first line of the Fig.~\ref{mainrule}.
After that, one can understand at what angles $\beta$ of crystalline lattice rotation nonlinear circular dichroism appears:}
\noindent\rule{\linewidth}{0.4pt}

{If there exists a number  $\nu \in \mathbb{Z}$ such that 
the difference between the angular momentum projections $\Delta m_\chi$ coming from the nonlinear susceptibility tensor is equal to $\nu\mathfrak{n}$, where $\mathfrak{n}$ is related to the symmetry group,  we introduce the number $s$:
\begin{equation}
   s = \Delta m_\chi = \nu \mathfrak{n}. 
\end{equation}
The nonlinear circular dichroism in such a nanoparticle appears for any angles of crystalline lattice rotation
\begin{equation}
   \beta \neq \frac{\pi \nu}{s}. \label{pinu}
\end{equation}
If such integer $\nu \in \mathbb{Z}$ exists for several $\Delta m_\chi$, then it is necessary to choose angles $\beta \neq \pi \nu/s$ that are common to all $s = \Delta m_\chi$. }
\noindent\rule{\linewidth}{0.4pt}

In short words, CD is present, when the modes, excited by the different terms of $\hat\chi^{(2)}$ tensor have the same symmetry and interfere, while the phases depend on the incident polarization. And this is possible only if \eqref{pinu} is satisfied. The application of the rule is shown in Fig.~\ref{mainrule}. We did not include the trivial case of the crystalline lattice BaTiO$_3$, [001]$\|z$.  For this case we only have single $m_\chi=0$. Therefore the circular dichroism in such a crystalline lattice is not possible for any symmetry of the nanoparticle, while nothing depends on the incident polarization. \blue{In this paper, we omit the discussion of SH-CD strength for every particular case, while the huge values close to $1$ were already achieved~\cite{Frizyuk2021-NonlinearCircularDi}, and an example is given in the Suppl. info. However, we should note, that SH-CD will possess local maxima close to the resonances of the modes with ${m^{2\omega}+\Delta m_\chi}$, while the phases changing rapidly allowing constructive or destructive interference.}
 
The correctness of the results was verified with the symmetry analysis of the overlap integrals (see Supplementary info) and numerical modelling in COMSOL Multiphysics\texttrademark~\blue{in the undepleted pump approximation. The method is well-known, and described, for example, in~\cite{Carletti2019-Secondharmonicgener,Timofeeva2018-AnapolesinFree-Stan}}.

\paragraph*{Conclusion.} 
In our work, we provide the simple rule, based on angular momentum projection conservation and modal analysis,  which immediately says, when the nonlinear circular dichroism appears for all possible achiral  $\text{C}_{\mathfrak{n}\text{v}} (\text{D}_{\mathfrak{n} \text{h}})$ nanoparticle's shapes and crystalline lattices. \blue{Moreover, the main formula does not depend on the angular momentum of the incident wave, which suggests that it could be applied to irradiation with beams of arbitrary angular momentum. The method can also provide ideas on how to describe the appearance of CD in other systems where the lattice and metaatom symmetries interact, for example, in metasurfaces where the metaatoms are rotated. We can also assume that the oblique wave incidence can be described as a normal incidence on a rotated particle, which can be equivalent to a particle of lower symmetry. A detailed study of these hypotheses may be the subject of further research.}\\

Authors acknowledge prof. Yuri Kivshar for the valuable advises and the initial impetus for this work, Mikhail Glazov and other members of Low-dimensional seminar in Ioffe Institute, Mihail Petrov, Ivan Toftul, and Kirill Koshelev for many fruitful discussions. 
This work was supported by
Russian Science Foundation (project 22-12-00204) and Priority 2030 Federal Academic Leadership Program. KF
acknowledges support from the Foundation for the Advancement of Theoretical Physics and Mathematics
“BASIS” (Russia)
{\paragraph*{Data availability.} Data and codes for the results presented in this paper are available in Supplementary information and Ref.~\cite{gitt}}
\bibliography{sample}
\newpage

\end{document}


\preprint{APS/123-QED}

\title{When does nonlinear circular dichroism appear in achiral dielectric nanoparticles? Supplementary information}

\author{Anastasia Nikitina}
\email{anastasia.nikitina@metalab.ifmo.ru}
\author{Anna Nikolaeva}%
\author{Kristina Frizyuk}%
 \email{k.frizyuk@metalab.ifmo.ru}
\affiliation{%
 The School of Physics and Engineering, ITMO University, Saint-Petersburg, Russia \\ 
}%

\date{\today}

\begin{abstract}
Here we provide an example of detailed theoretical description, which shows, how the main formula is obtained. We also show the nonlinear polarization for different cases. 
\end{abstract}

\maketitle



\onecolumngrid



\subsection{Theory} 
The SH polarization is written as follows~\cite{Boyd2003}:
\begin{align}
 \vec P^{2\omega} (\vec r)=\varepsilon_0 \hat\chi^{(2)} \vec E^\text{in} (\vec r)\vec E^\text{in} (\vec r)\label{pol},
\end{align}
where $\vec E^\text{in}(\vec r)$ is the fundamental field inside the nanoparticle, $\hat\chi^{(2)}$ --- nonlinear susceptibility tensor. 
The SH field can be written with help of the dyadic Green's function formalism~\cite{Frizyuk_Volkovskaya_Smirnova_Poddubny_Petrov_2019,Novotny}
\begin{align}
\mathbf{E}^{2\omega}({\mathbf{r}})&=(2\omega)^2\mu_0\int\limits_V  \dd V' \hat {{\bf G}}({\bf r,r'},k) \mathbf{P}^{2\omega} (\mathbf{r}^\prime)=\nonumber \\
&=(2\omega)^2\mu_0\int\limits_V \dd V' \sum_{j} \frac{\mathbf{E}_{j}(\mathbf{r}) \otimes \mathbf{E}_{j}\left(\mathbf{r}^{\prime}\right)}{2 k\left(k-k_{j}\right)} \mathbf{P}^{2\omega} (\mathbf{r}')=
\\
&=(2\omega)^2\mu_0\sum_{n}\frac{1}{2k\left(k-k_{j}\right)} \mathbf{E}_{j}(\mathbf{r})
\nonumber \int\limits_V \dd V'   \mathbf{E}_{j}\left(\mathbf{r}^{\prime}\right) \mathbf{P}^{2\omega} (\mathbf{r}'), \label{shfield}
\end{align}
here, we employ the resonant-state expansion of the Green function~\cite{PhysRevA.90.013834}, where $\mathbf{E}_{j}(\mathbf{r})$ is system's quasi-normal mode~\cite{Lalanne2018-LightInteractionwit,Gigli2020-Quasinormal-ModeNon} with corresponding frequency $\omega_j=ck_j$, the fundamental frequency is $\omega=ck$.  The integration is over the particle's volume. Note that here the many subtleties arising in the decomposition over resonance states do not play a role, nor does the fact that the Green's function is defined only in a restricted region. To explain the presence or absence of circular dichroism it is sufficient to know only the symmetric behavior of all fields. The key role in our results is played by the overlap integral over the particle's volume with the nonlinear polarization:
\begin{equation}
D_j=\int\limits_V \dd V' \mathbf{E}_{j} \left(\mathbf{r}^{\prime}\right) \mathbf{P}^{2\omega} (\mathbf{r}')\label{overlap}.
\end{equation}
The SH intensity will be proportional to the square of the absolute value of the coefficient $D_j$. 
We assume that each eigenmode consists of a infinite set of multipoles~\cite{Xiong2020-Ontheconstraintsof, Gladyshev_Frizyuk_Bogdanov_2020,PhysRevA.90.013834}:
\begin{align}
  \mathbf{E}_{n}(\mathbf{r})=\sum_{m,\ell}c_{m\ell}\mathbf{W}_{m\ell} (\mathbf{r}),
\end{align}
where $\mathbf{W}_{m\ell} (\mathbf{r}) $ is a complex vector spherical harmonic~\cite{Deriy2022-BoundStatesintheC, VMK, Bohren1998Mar}, where $\ell$ denote the total angular momentum, and $m$ --- the projection~\cite{Akhiezer}. For each mode, we will be only interested in the set of $m$s, which contribute into the mode.


\subsection{The nonlinear polarization}
Normally incident circularly polarized wave in the multipole expansion contains only vector spherical harmonics $\mathbf{W}_{m\ell} (\mathbf{r})$ with $m = \pm 1$ and is written as following:
\begin{align}
    \mathbf{E}^\text{wave} = E_x (\mathbf{e}_r + \iu \mathbf{e}_\varphi) \eu^{\pm \iu \varphi}.
\end{align}
{To figure out which eigenmodes are excited in the nanostructure with symmetry group $\text{C}_{\mathfrak{n}\text{v}} (\text{D}_{\mathfrak{n} \text{h}})/\text{C}_{\mathfrak{n}\text{h}}$, one can use the rule of conservation of the angular momentum projection. Multipole expansion of the field inside the nanoparticle should include vector spherical harmonics only with $m = \pm 1 + \mathfrak{n}\nu$, where $\nu \in \mathbb{Z}$, and $\mathfrak{n}$ is related to the symmetry group $\text{C}_{\mathfrak{n}\text{v}} (\text{D}_{\mathfrak{n} \text{h}})$. One can verify this with the help of multipole decomposition of the modes~\cite{Gladyshev_Frizyuk_Bogdanov_2020}. In this case, the field can be written as following:}
\begin{align}
    \mathbf{E}^\text{in}(r,z,\varphi) = \sum\limits_{\nu} \mathbf{E}_{\nu}(r,z) \eu^{\pm 1\iu\varphi + \mathfrak{n}\nu \iu\varphi}. \label{einc}
\end{align}
Using the method, given in~\cite{Frizyuk2021-NonlinearCircularDi}, one can obtain the $\hat\chi^{(2)}$ tensor in cylindrical coordinates, and then the nonlinear polarization. For example, the $\hat{\chi}^{(2)}$ tensor for GaAs[001]\texorpdfstring{$\|$}{Lg}z has only one independent component $\chi_{xyz} = \chi_{yxz} = \chi_{xzy} = \chi_{yzx} = \chi_{zxy} = \chi_{zyx} = \chi$. It can be written as following:
\begin{align}
    \hat{\chi}_{ijk} = \chi_{ijk}\vec {e}_{i}\vec {e}_{j}\vec {e}_{k}.
\end{align}
Using the relations for the cylindrical coordinate system
\begin{align}
    &\vec {e}_{x} = \cos\varphi \vec {e}_{r} - \sin\varphi \vec {e}_{\varphi}\nonumber\\
    &\vec {e}_{y} = \sin\varphi \vec {e}_{r} + \cos\varphi \vec {e}_{\varphi}\\   
    &\vec {e}_{z} = \vec {e}_{z},\nonumber
\end{align}
we can rewrite the $\hat{\chi}^{(2)}$ tensor in cylindrical coordinates. For example, one can consider sum of two components:
\begin{eqnarray}
\vec {e}_{z}\vec {e}_{y}\vec {e}_{x}+\vec {e}_{z}\vec {e}_{x}\vec {e}_{y} &= &
{\eu^{2\iu\varphi}}\left(\frac1{2\iu}\vec {e}_{z}\vec {e}_{r}\vec {e}_{r}-\frac1{2\iu}\vec {e}_{z}\vec {e}_{\varphi}\vec {e}_{\varphi}+
\frac1{2}\vec {e}_{z}\vec {e}_{\varphi}\vec {e}_{r}+\frac1{2}\vec {e}_{z}\vec {e}_{r}\vec {e}_{\varphi}\right) + 
\\  &&+
{\eu^{-2\iu\varphi}}\left(-\frac1{2\iu}\vec {e}_{z}\vec {e}_{r}\vec {e}_{r}+\frac1{2\iu}\vec {e}_{z}\vec {e}_{\varphi}\vec {e}_{\varphi}+
\frac1{2}\vec {e}_{z}\vec {e}_{\varphi}\vec {e}_{r}+\frac1{2}\vec {e}_{z}\vec {e}_{r}\vec {e}_{\varphi}\right). \nonumber
\end{eqnarray}
Other components can be written in the same way. It is worth noting, that the $\hat\chi^{(2)}$ tensor depends on $\varphi$, but contains only two terms with $m = \pm 2$. We are only interested in the symmetry behavior of the tensor under rotations around the $z$-axis, but not in its specific form. Thus, knowing the behavior of the basis vectors, one can show that $\hat\chi^{(2)}$ transforms for GaAs[001]\texorpdfstring{$\|$}{Lg}z  as $\sin[2\varphi]$. In addition, we should mention that crystalline lattice can be rotated by some angle $\beta$. In this case, the tensor transforms as $\sin[2(\varphi-\beta)]$.

\begin{figure}[ht!]
\centering
\includegraphics[width=\linewidth]{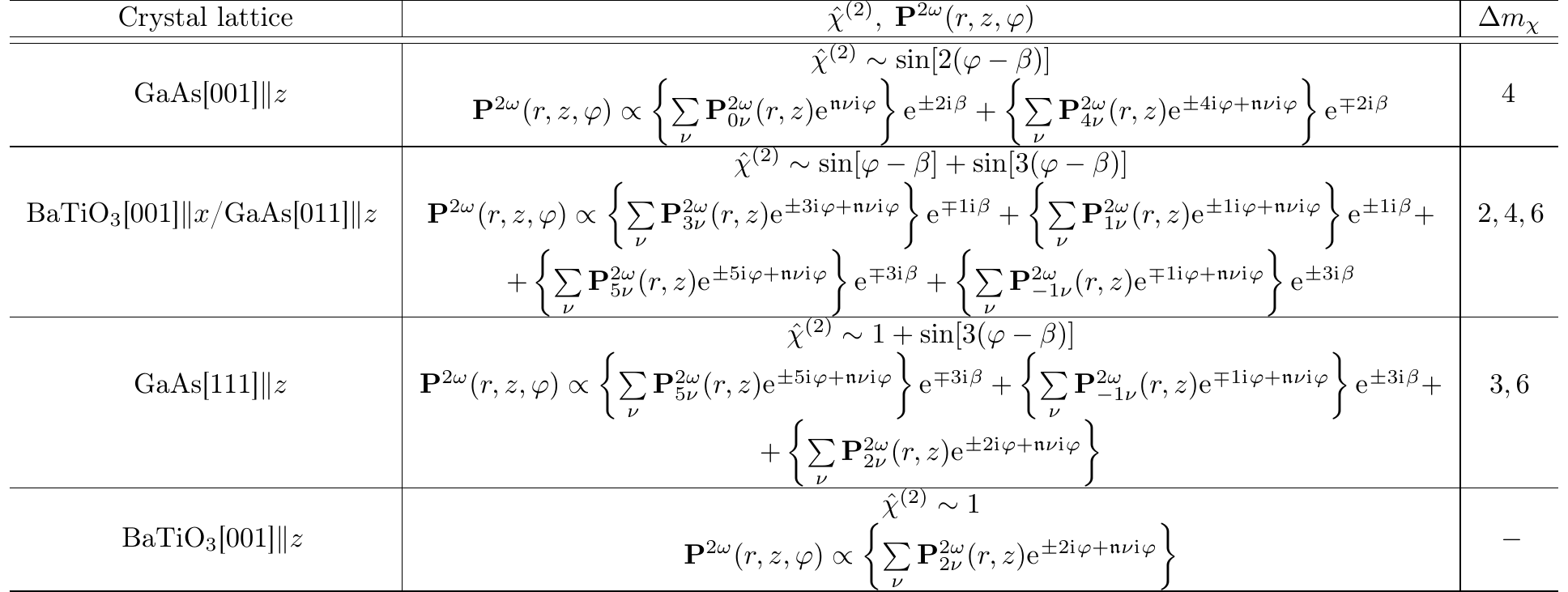}
\caption{The  symmetry behavior of the susceptibility tensor $\hat\chi^{(2)}$ under rotations around the $z$-axis and nonlinear polarization $\mathbf{P}^{2\omega}(r,z,\varphi)$ in cylindrical coordinates for each crystalline lattice. In the third column we also show $\Delta m_\chi $ --- difference between the angular momentum projections in tensor $\hat\chi^{(2)}$.}
\label{polandmomentum}
\end{figure}

Similarly, we defined the symmetry behavior of the $\hat\chi^{(2)}$ tensor under rotations around the $z$-axis for each crystalline lattice, that is shown in the Fig.~\ref{polandmomentum}. 
After that, with the help of~\eqref{pol} and \eqref{einc}, we calculated SH polarization for each case (See Fig.~\ref{polandmomentum}). 
To obtain, whether there SH-CD exists, we can further substitute the nonlinear polarization into the Green's function, analyse the symmetry of all the modes excited, and compare the intensity under both polarizations for each case. {However, instead of the sophisticated calculations, one can note some patterns, and use the short rule that we have formulated and explained below.}
\subsection{The conditions for obtaining the SH-CD}
First of all, let us note that the nonlinear polarization contains several terms, each of them possess a phase multiplier, which depends on $\beta$ multiplied by $m_\chi$, which comes from $\hat\chi^{(2)}$ in cylindrical coordinates.
For each term in $\vec P^{2\omega}$, the conservation of the momentum projection is fulfilled, indeed, from the incident wave $\vec E^\text{in}\vec E^\text{in}$ comes $\pm2+\mathfrak{n}\nu$, and additional $m_\chi$ comes from the crystalline lattice. For example, for the GaAs, $[001]||z$ case,  $m_\chi=\pm2$, for $[111]||z$ $m_\chi=0, \pm3$, and so on. One can easily extract these numbers, rewriting tensors in cylindrical coordinates. 

After all, the nonlinear polarization excites the modes of a nanostructure on the $2\omega$ frequency. 
The most important condition for the existence of the SH-CD:
{\it{at least two different terms of the polarization should excite the same mode}} (or several different modes with the same symmetry). Let us consider the case of GaAs, $[001]||z$, D$_{2\text{h}}$(C$_{2\text{v}}$) symmetry of the nanostructure.
In this case, the overall integral intensity $I^{2\omega, \text{total}}$, which is for a particular mode proportional to $|D_n|^2$:
\begin{gather} \label{difd}
   I^{2\omega, \text{total}}\propto |D_n|^2= |a \eu^{\pm2 i \beta}+b\eu^{\iu\alpha} \eu^{\mp 2 \iu \beta}| =a^2+b^2+2ab\cos(\pm4\beta+\alpha), 
\end{gather}
where $a$ and $b$ can be assumed real due to the phase $\alpha$ between the two terms in $D_n$ coefficient, coming from different terms in polarization. Phase $\alpha$ quickly varies  in the vicinity of resonances, so the local maxima of the SH-CD are obtained. In case if all terms of the polarization excite modes of the different symmetry, each $D_n$-coefficient contains only one term, so the $a^2+b^2+2ab\cos(\pm4\beta+\alpha)$ terms, responsible for the CD as in \eqref{difd} are never obtained, while the modes do not interfere. 
{To satisfy the condition, it is necessary that  the difference $\Delta m_\chi$ between the angular momentum coming from tensor, related to different terms $\mathbf{P}^{2\omega} (\mathbf{r})$, is equal to a multiple of the rotational symmetry of the nanostructure, in other words, the number $s$ should exist:}
\begin{align}
   s = \Delta m_\chi = \mathfrak{n}\nu, \; \nu \in \mathbb{Z},
    \label{deltam}
\end{align}
{where $\Delta m_\chi$ --- difference between the angular momentum projections in tensor $\hat\chi^{(2)}$. For one crystalline lattice can be several $\Delta m_\chi$. 
In Fig.~\ref{polandmomentum} all possible $\Delta m_\chi$ for several crystalline lattices are shown. It is worth noting, that always $\Delta m_\chi\leq 6$, while it can't be greater for a third-rank tensor. Thus if $\mathfrak{n}\geq 7$, the circular dichroism is impossible for second harmonic. 
The fulfillment of \eqref{deltam} provides the possibility to obtain the SH-CD for some angles $\beta$.
However it is still not enough for the existence of nonlinear circular dichroism, because the interference contribution does not depend on polarization for some crystalline lattice orientations, i.e for some angles $\beta$ included in the expression for  $I^{2\omega,\text{total}}$. Indeed, analogously to the \eqref{difd}, we will obtain the expressions of the following form:
\begin{gather} \label{difdd}
   I^{2\omega, \text{total}}\propto a^2+b^2+2ab\cos(\pm s\beta+\alpha).
\end{gather}
 This leads to the necessary condition for the orientation of the crystalline lattice: if \eqref{deltam} true and $s$ exists, and angles of crystalline lattice rotation $\beta$ do not equal to $\pi \nu / s,\; \nu \in \mathbb{Z}$, then and only then SH-CD exists.}
\subsection{Numerical modelling in COMSOl Multiphysics\texttrademark}

\begin{figure}[ht!]
\centering
\includegraphics[width=1.03\linewidth]{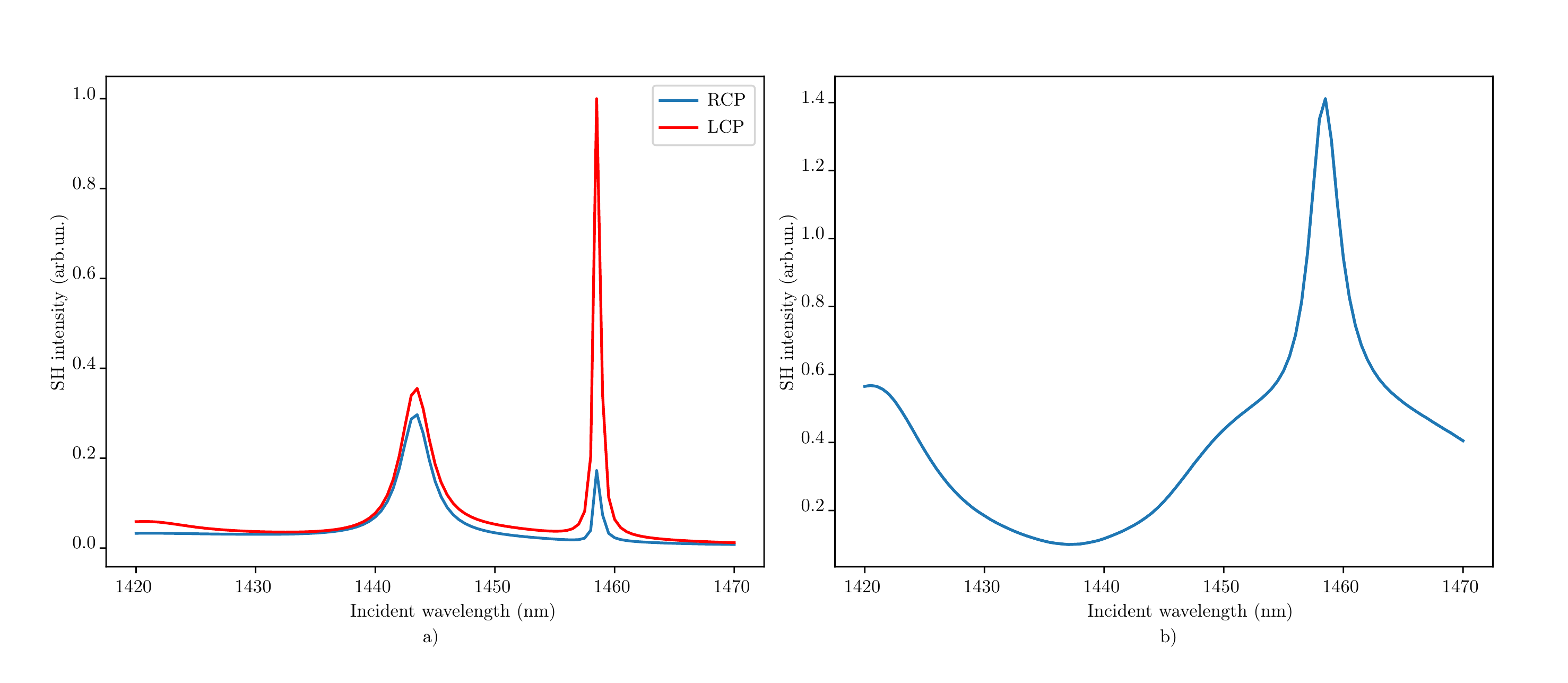}
\caption{\blue{a) Numerically calculated SH intensity in dimer with BaTiO$_3$, [001]$\|x$ $\hat\chi^{(2)}$ tensor for both RCP and LCP polarization of incident wave, and b) the resulting nonlinear
circular dichroism given by formula $\text{SH-CD}= 2(I_\text{RCP}^{2\omega}-I_\text{LCP}^{2\omega})/(I_\text{RCP}^{2\omega}+I_\text{LCP}^{2\omega})$ of the SH signal.}}
\label{dichroismCmsol}
\end{figure}

\blue{Our results were verified with numerical calculations  of the SH intensity in COMSOL Multiphysics\texttrademark \ for all symmetries and crystal lattices considered in the paper. We show one illustrative case, in Fig.~\ref{dichroismCmsol}. The BaTiO$_3$~\cite{Cabuk2012-Thenonlinearoptical} dimer, [001]$\|x$, is expected to show larger numbers of dichroism, while it vanishes only for $\beta=\pi\nu/2$ rotation angles. Parameters of the system we used: diameter and height of single cylinder $d = 550 $ nm and $h = 450$ nm, and distance between the cylinders $l = 30$ nm. We took the constant refractive index $n_{\text {refr}}=3$ in the range 1420-1470 nm, which is close to the values of a typical dielectric material. The cylinders' size is quite big to obtain the higher-order multipolar resonances, $m\sim 5$. In the vicinity of these resonances, huge values of SH-CD were obtained. Generally, the basic recipe to obtain huge values of SH-CD is to take quite big structures, where the multipolar resonanсes with $m$ from different polarization terms (see Table \ref{polandmomentum}) are presented.}

\section{MatLab code for tensor rotation}

In general case, it can be useful to consider an arbitrary orientation of a crystalline lattice, while in literature only a specific one is given. 
We provide the MatLab code to obtain the $\hat\chi^{(2)}$ components for a rotated lattice~\cite{gitt}. 
General formula is
\begin{equation}
\chi^{(2)}_{(i j k)_{\text{rotated}}}= M_{i a} M_{j b} M_{k c} \chi^{(2)}_{(a b c)_{\text{underrotated}}}
\end{equation}
Rotation matrix $\hat{M}$ is defined as
\begin{equation}
\mathbf{A}^{\text{rotated}} = \hat{M}^{-1} \mathbf{A}^{\text{underrotated}}
\end{equation}
The matrix of a proper rotation $\hat{M}$ by angle $\beta$ around the axis $r=(x, y, z)$, a unit vector with $x^2+y^2+z^2=1$

\begin{gather}
{M={\begin{bmatrix} \cos \beta +x^{2}\left(1-\cos \beta \right) & xy\left(1-\cos \beta \right)-z\sin \beta & xz\left(1-\cos \beta \right)+y\sin \beta \\ yx\left(1-\cos \beta \right)+z\sin \beta & \cos \beta +y^{2}\left(1-\cos \beta \right)& yz\left(1-\cos \beta \right)-x\sin \beta \\
zx\left(1-\cos \beta \right)-y\sin \beta & zy\left(1-\cos \beta \right)+x\sin \beta & \cos \beta +z^{2}\left(1-\cos \beta \right)\end{bmatrix}}}
\end{gather}

\subsection{ Code structure }
Main.m --- main matlab file, which rotates $3\times 6$ matrix across  $(x,y,z)$ axis, where $x^2+y^2+z^2=1$ by the $\beta$ angle. The result is rotated matrix in $3\times6$ form.

\begin{lstlisting}
% Write your tensor in 3*6 form
dd=[[0,0,0,0,17,0];[0,0,0,17,0,0];[15.7,15.7,6.8,0,0,0]]%batio3 example

% Converting into 3*3*3 matrix, using the symmetry on the last two indices
RR = zeros(3, 3, 3);
RR(:,:,1)=[[dd(1,1),dd(1,6),dd(1,5)];[dd(2,1),dd(2,6),dd(2,5)];[dd(3,1),dd(3,6),dd(3,5)]];
RR(:,:,2)=[[dd(1,6),dd(1,2),dd(1,4)];[dd(2,6),dd(2,2),dd(2,4)];[dd(3,6),dd(3,2),dd(3,4)]];
RR(:,:,3)=[[dd(1,5),dd(1,4),dd(1,3)];[dd(2,5),dd(2,4),dd(2,3)];[dd(3,5),dd(3,4),dd(3,3)]];

% Rotaton. Here put axes of rotation in form (x,y,z), where x^2+y^2+z^2=1
x=0;
y=1;
z=0;

% Here put rotation angle
beta=pi/2;

% XX - Rotated matrix in form 3*3*3
XX=Rotation_tensor(x,y,z,beta,RR);

%Converting  into 3*6 form
dd_new=[[0,0,0,0,0,0];[0,0,0,0,0,0];[0,0,0,0,0,0]];
dd_new(1,1)=XX(1,1,1);
dd_new(1,2)=XX(1,2,2);
dd_new(1,3)=XX(1,3,3);
dd_new(1,4)=XX(1,3,2);
dd_new(1,5)=XX(1,3,1);
dd_new(1,6)=XX(1,2,1);

dd_new(2,1)=XX(2,1,1);
dd_new(2,2)=XX(2,2,2);
dd_new(2,3)=XX(2,3,3);
dd_new(2,4)=XX(2,3,2);
dd_new(2,5)=XX(2,3,1);
dd_new(2,6)=XX(2,2,1);

dd_new(3,1)=XX(3,1,1);
dd_new(3,2)=XX(3,2,2);
dd_new(3,3)=XX(3,3,3);
dd_new(3,4)=XX(3,3,2);
dd_new(3,5)=XX(3,3,1);
dd_new(3,6)=XX(3,2,1);
dd_new
%%
\end{lstlisting}

Mmatrix.m --- function, which takes parameters of rotation $(x,y,z,\beta,$ where $x^2+y^2+z^2=1)$ and build rotation matrix $\hat{M}$.

\begin{lstlisting}


% Rotation matrix (x,y,z-axis, beta-angle of rotation)
function MM=Mmatrix(x,y,z,beta) 
MM=zeros(3,3);

MM(1,1)=cos(beta)+(1-cos(beta))*x.^2;
MM(1,2)=(1-cos(beta))*x*y-sin(beta)*z;
MM(1,3)=(1-cos(beta))*x*z+sin(beta)*y;

MM(2,1)=(1-cos(beta))*x*y+sin(beta)*z;
MM(2,2)=cos(beta)+(1-cos(beta))*y.^2;
MM(2,3)=(1-cos(beta))*y*z-sin(beta)*x;

MM(3,1)=(1-cos(beta))*x*z-sin(beta)*y;
MM(3,2)=(1-cos(beta))*z*y+sin(beta)*x;
MM(3,3)=cos(beta)+(1-cos(beta))*z.^2;


return
\end{lstlisting}

Rotation\_tensor.m --- function, which takes parameters of rotation $(x,y,z,\beta,$ where $x^2+y^2+z^2=1)$ and initial matrix $RR$ ($ 3\times 3\times 3$). The result is rotated $RR$ matrix in $3\times 3\times 3$ form.

\begin{lstlisting}
% RR - initial tensor
% (x,y,z) - the axis around which the rotation through the angle beta, x^2+y^2+z^2=1
% XX - rotated tensor
function XX=Rotation_tensor(x,y,z,beta,RR)
MM=Mmatrix(x,y,z,beta); % rotation matrix, Mmatrix - rotation function
MM1=inv(MM);
XX=zeros(3, 3, 3);
for i=1:3
    for j=1:3
        for k=1:3
for a=1:3
    for b=1:3
        for c=1:3    
            XX(i,j,k)=XX(i,j,k)+MM(k,c)*MM(j,b)*MM(i,a)*RR(a,b,c);
        end
    end
end
        end
    end
end
return
\end{lstlisting}

\bibliography{sample}